\documentstyle[aasms4]{article}

\begin{document}

\title{RELATIVISTIC THERMAL BREMSSTRAHLUNG GAUNT 
FACTOR FOR THE INTRACLUSTER PLASMA. II. ANALYTIC FITTING FORMULAE}

\author{NAOKI ITOH, TSUYOSHI SAKAMOTO, AND SHUGO KUSANO,}
\affil{Department of Physics, Sophia University, 7-1 Kioi-cho, Chiyoda-ku, Tokyo, 102-8554, Japan;}
\affil{n\_itoh, t-sakamo, s-kusano@hoffman.cc.sophia.ac.jp}

\author{SATOSHI NOZAWA}

\affil{Josai Junior College for Women, 1-1 Keyakidai, Sakado-shi, Saitama, 350-0290, Japan;}

\affil{snozawa@galaxy.josai.ac.jp}



\centerline{AND}

\author{YASUHARU KOHYAMA}

\affil{Fuji Research Institute Corporation, 2-3 Kanda-Nishiki-cho, Chiyoda-ku, Tokyo, 101-8443, Japan;}
\affil{kohyama@star.fuji-ric.co.jp}


\begin{abstract}

  We present accurate analytic fitting formulae which summarize the results of the recent calculation by Nozawa, Itoh \& Kohyama on the relativistic thermal bremsstrahlung Gaunt factor for the intracluster plasma. The fitting is carried out for $1 \leq Z_j \leq 28$, $6.0 \leq \log T \leq 8.5$, $-4.0 \leq \log (\hbar \omega / k_BT) \leq 1.0$ by combining the relativisitic Gaunt factor with the nonrelativistic exact Gaunt factor smoothly, where $Z_j$ is the charge number of the ion, $T$ is the temperature in kelvins, and $\omega$ is the angular frequency of the emitted photon.  We also present an accurate analytic fitting formula for the nonrelativistic exact Gaunt factor.  This fitting formula is best suited for relatively low-temperature plasmas in which the relativisitic effect is negligible.  The present analytic fitting formulae will be useful for the analysis of the X-ray emission which comes from the intracluster plasma as well as the other X-ray sources. 

\end{abstract}


\section{INTRODUCTION}

  Three of the present authors (S.N., N.I., and Y.K.) have recently carried out
 accurate calculations on the relativistic thermal bremsstrahlung Gaunt factor 
for the intracluster plasma (Nozawa, Itoh, \& Kohyama 1998). Their calculation 
is based on the method of Itoh and his collaborators (Itoh, Nakagawa, 
\& Kohyama 1985; Nakagawa, Kohyama, \& Itoh 1987; Itoh, Kojo, \& Nakagawa 1990;
 Itoh et al. 1991,1997). In calculating the relativistic thermal bremsstrahlung
 Gaunt factor for the high-temperature, low-density plasma, Nozawa, Itoh, \& 
Kohyama (1998) have made use of the Bethe-Heitler cross section (Bethe \& 
Heitler 1934) corrected by the Elwert factor (Elwert 1939). They have also 
calculated the Gaunt factor using the Coulomb-distorted wave functions for nonrelativistic electrons following the method of Karzas \& Latter (1961). They have presented the results of the numerical calculations in the form of extensive numerical tables. The grid of their numerical table is made fine enough so that
 smooth interpolations can be carried out for the applications of the results 
to the analyses of various  astrophysical observational data.

  Other references on the calculation of thermal bremsstrahlung Gaunt factor include Culhane (1969), Culhane \& Acton (1970), Raymond \& Smith (1977), Gronenschild \& Mewe (1978), Mewe, Lemen, \& van den Oord (1986), and Carson (1988).

  In this  paper we will present accurate analytic fitting formulae which 
summarize the results of the extensive numerical tables presented by Nozawa, Itoh, \& Kohyama (1998).  Analytic fitting formulae can be readily implemented in the computer programs for the analyses of the observational data. Therefore, we recommend the present analytic fitting formulae be used widely by the observers for the analyses of their observational data.

  The present paper is organized as follows. We will give the analytic fitting formulae for the relativistic Gaunt factor in $\S$2.  We will give the analytic formula for the nonrelativistic exact Gaunt factor in $\S$3.  Concluding remarks will be given in $\S$4. 

\section{ANALYTIC FITTING FORMULAE FOR THE RELATIVISTIC GAUNT FACTOR}

  The thermal bremsstrahlung emissivity is expressed in terms of the thermally averaged relativistic Gaunt factor $g_{Z_j}$ (Nozawa, Itoh, \& Kohyama 1998) by
\begin{eqnarray}
< W(\omega) > d \omega  & = & 1.426 \times 10^{-27}g_{Z_j}(T,u)
\, n_{e} n_{j} Z_{j}^{2} T^{1/2} \nonumber  \\
& \times & e^{-u} \, du  \, \hspace{0cm} \, {\rm ergs \, \, s^{-1} \, cm^{-3}} 
 \,  ,  \\
u & \equiv & \frac{\hbar \omega}{k_{B}T}  \, ,
\end{eqnarray}
where $\omega$ is the angular frequency of the emitted photon, $T$ is the temperature of the electrons (in kelvins), $n_e$ is the number density of the electrons (in cm$^{-3}$), and $n_j$ is the number density of the ions with the charge $Z_j$ (in cm$^{-3}$).  In Nozawa, Itoh, \& Kohyama (1998), the nonrelativistic Gaunt factor $g_{NR}$ has been also calculated. This  Gaunt 
factor is exact in the low-temperature limit.  At intermediate temperatures, it has been confirmed by Nozawa, Itoh, \& Kohyama (1998) that the relativistic 
Gaunt factor which has been calculated with the use of the Bethe-Heitler cross section corrected by the Elwert factor shows excellent agreement with the nonrelativistic exact Gaunt factor. At higher temperatures, the nonrelativistic Gaunt factor deviates from the relativistic Gaunt factor because of the insufficiency of the nonrelativistic approximation. These two Gaunt factors 
have been tabulated in Nozawa, Itoh, \& Kohyama (1998).

  We have thoroughly examined the numerical results presented in 
the tables in Nozawa, Itoh, \& Kohyama (1998). We have confirmed that the data are almost perfect. Nevertheless, we have found 11 cases where the numerical 
computation has not converged sufficiently.  All are for the nonrelativistic exact Gaunt factor at low temperatures. Therefore, they will not be relevant to the intracluster plasma. However, for the sake of completeness, we will replace
some of the tables in Nozawa, Itoh,  \& Kohyama (1998) with the revised ones. In particular, we replace the tables on pages 541, 551, 552 of Nozawa, Itoh,  \& Kohyama (1998) with TABLES 1-3.  In TABLES 1-3, $\gamma^{2}$ is defined by
\begin{eqnarray}
\gamma^{2} & = & \frac{Z_{j}^{2} {\rm R_{y}}}{k_{B}T}  \, = \, Z_{j}^{2} \, \frac{1.579 \times 10^{5} {\rm K}}{T} \, .
\end{eqnarray}

  At sufficiently high temperatures, we adopt the relativistic Gaunt factor. At sufficiently low temperatures, we adopt the nonrelativistic exact Gaunt factor.
At intermediate temperatures, these two Gaunt factors coincide with each other for small values of $Z_j$. For larger values of $Z_j$, the two Gaunt factors show small discrepancies even at intermediate temperatures.  Therefore, we generally  interpolate between the two Gaunt factors smoothly at intermediate temperatures. To be more precise, we find the point at which the discrepancy between the two Gaunt factors (for fixed values of $Z_j$ and {\it u}) is the smallest as a function of the temperature. Then we interpolate between the two Gaunt factors
smoothly using a sine function. The temperature range for the interpolation is $\Delta$ $\log T$ = $\pm$ 0.1 to $\pm$ 0.5 with respect to the central temperature at which the discrepancy is the smallest depending on the minimum value of the discrepancy.

  We give analytic fitting formulae for $Z_j = 1 - 28$ . The range of the fitting is $6.0 \leq \log T \leq 8.5$, $-4.0 \leq \log u \leq 1.0$.  We express the Gaunt factor by
\begin{eqnarray}
g_{Z_{j}} & = &  \sum^{10}_{i,j = 0} a_{ij}t^i U^j  \,  ,  \\
t & \equiv & \frac{1}{1.25}[\log T - 7.25] \,  ,  \\
U & \equiv & \frac{1}{2.5}[\log u + 1.5] .   
\end{eqnarray}
The coefficients $a_{ij}$ for $Z_{j} = 1 - 28$ are presented in TABLE 4.
The accuracy of the fitting is generally better than 0.1\%.

\section{ANALYTIC FITTING FORMULA FOR THE NONRELATIVISTIC EXACT GAUNT FACTOR}

  The thermal bremsstrahlung emissivity in the nonrelativistic limit is expressed in terms of the nonrelativistic exact Gaunt factor $g_{\rm NR}$ (Nozawa, Itoh, \& Kohyama 1998) by 
\begin{eqnarray}
< W(\omega) >_{\rm NR} d \omega & = & 1.426 \times 10^{-27}g_{\rm NR}( \gamma^2 ,u) \, n_{e} n_{j} Z_{j}^{2} T^{1/2} \nonumber  \\
& \times & e^{-u} \ du \hspace{0cm} \ {\rm ergs\,\, s^{-1} \, cm^{-3}} \, , \\
u        & \equiv & \frac{\hbar \omega}{k_{B}T}  \, , \\
\gamma^2 & \equiv & \frac{{Z_j}^2{\rm Ry}}{k_B T} = {Z_j}^2 \frac{1.579 \times 10^5{\rm K}}{T} \, ,
\end{eqnarray}
where $\omega$ is the angular frequency of the emitted photon, $T$ is the temperature of the electrons (in kelvins), $n_e$ is the number density of the electrons (in cm$^{-3}$), and $n_j$ is the number density of the ions with the charge $Z_j$ (in cm$^{-3}$).  It should be noted that the thermal bremsstrahlung emissivity in the nonrelativistic limit is a function of $\gamma^2$ and $u$ only. It does not depend on $Z_j$ and $T$ separately, but on the ratio ${Z_j}^2 / T $. This is a remarkable fact for nonrelativistic electrons.

  In Figure 1 we show the nonrelativistic Gaunt factor as a function of $u$ for various values of $\gamma^2$. In Figure 2 we show the nonrelativistic Gaunt factor as a function of $\gamma^2$ for various values of $u$.

  We give an analytic fitting formula for the nonrelativistic exact Gaunt factor. The range of the fitting is $-3.0 \leq \log \gamma^2 \leq 2.0$, $-4.0 \leq \log u \leq 1.0$. We express the Gaunt factor by 
\begin{eqnarray}
g_{\rm NR} & =      &  \sum^{10}_{i,j = 0} b_{ij}\Gamma^i U^j  \,  ,  \\
\Gamma & \equiv & \frac{1}{2.5}[\log \gamma^2 + 0.5] \,  ,  \\
U      & \equiv & \frac{1}{2.5}[\log u + 1.5] .
\end{eqnarray}
The coefficients $b_{ij}$ are presented in TABLE 5.  The accuracy of the fitting is generally better than 0.1\%.

\section{CONCLUDING REMARKS}

  We have presented accurate analytic fitting formulae for the relativistic Gaunt factor as well as for the nonrelativistic exact Gaunt factor for thermal bremsstrahlung. The analytic fitting formulae have been constructed to reproduce the numerical results of the calculation reported in Nozawa, Itoh, \& Kohyama (1998).  The accuracy of the fitting is generally better than 0.1\%. The present fitting formulae can be used widely for the analysis of thermal bremsstrahlung radiation. 

  We thank Professor Y. Oyanagi for allowing us to use the least square
fitting program SALS.  We also wish to thank our referee for useful comments.  We are indebted to the editors Dr. Helmut A. Abt and Dr. F. W. Stecker for their valuable advices on the form of the paper.  This work is financially supported in part by the Grant-in-Aid of Japanese Ministry of Education, Science, Sports, and Culture under the contract \#10640289.

\newpage

\begin{center}
{\bf Figure Legends} \\
\end{center}
\begin{tabbing}
FIG.1. \= Nonrelativistic exact Gaunt factor as a function of $u$ for various values of $\gamma^2$. \\
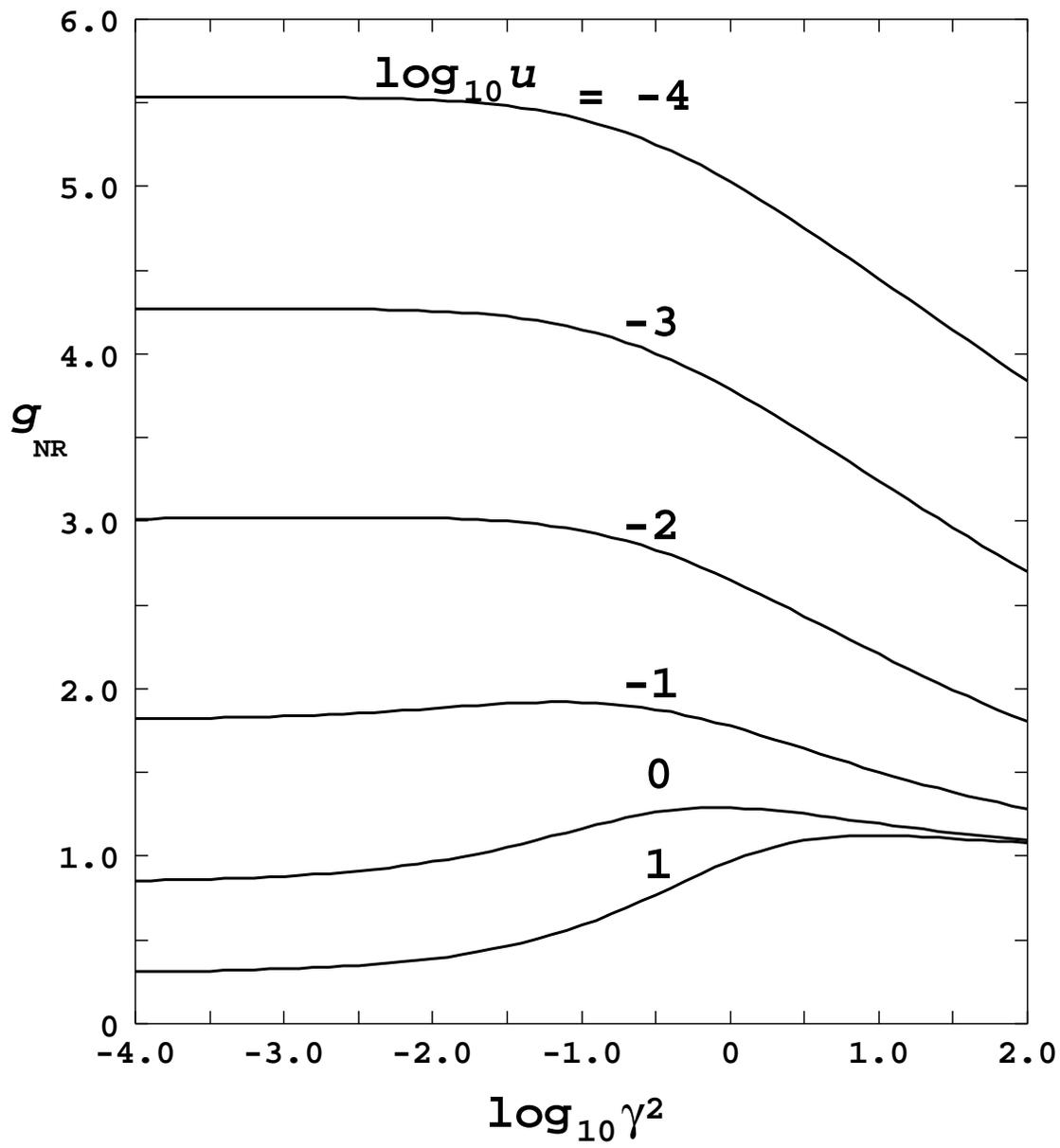
FIG.2. \= Nonrelativistic exact Gaunt factor as a function of $\gamma^2$ for various values of $u$.
\end{tabbing}

\end{document}